# The Emergence of Superconductivity in Heavy Electron Materials

Yi-feng Yang[1,2,*] & David Pines[3,4]

[1]Beijing National Laboratory for Condensed Matter Physics and Institute of Physics, Chinese Academy of Sciences, Beijing 100190, China

[2]Collaborative Innovation Center of Quantum Matter, Beijing 100190, China

[3]Department of Physics, University of California, Davis, CA 95616, USA

[4]Santa Fe Institute, Santa Fe, NM 87501, USA

[*]E-mail: yifeng@iphy.ac.cn

**Although the pairing glue for the attractive quasiparticle interaction responsible for unconventional superconductivity in heavy electron materials has been identified as the spin fluctuations that arise from their proximity to a magnetic quantum critical point[1-5], there has been no model to describe their superconducting transition at $T_c$ that is comparable to that found by Bardeen, Cooper, and Schrieffer[6] (BCS) for conventional superconductors where phonons provide the pairing glue. Here we propose a phenomenological BCS-like expression for $T_c$ in heavy electron materials, that is based on the unusual properties of the heavy electron normal state from which superconductivity emerges[7-12], and a simple model for the effective range and strength of the spin-fluctuation-induced quasiparticle interaction[13-17]. We show that it provides both a physical explanation and the first quantitative understanding of the pressure-induced variation of $T_c$ in the "hydrogen atoms" of unconventional superconductivity, $CeCoIn_5$ and $CeRhIn_5$ [18-23], and predicts scaling behavior and a dome-like structure for $T_c$ in all heavy electron quantum critical superconductors[24-30].**

In seeking to explain heavy electron superconductivity, it is helpful to begin by recalling the principal features of its remarkably similar emergence in two of the best-studied materials[18-23], $CeCoIn_5$ and $CeRhIn_5$. As may be seen in Fig. 1, there are three distinct regions of emergent heavy electron superconductivity in their pressure-temperature phase diagram that are defined by two ordering temperatures: the delocalization temperature, $T_L$, at which the collective hybridization of the local moments becomes complete; and the Néel temperature, $T_N$, that marks the onset of long-range antiferromagnetic order of the hybridized local moments.

Region I: $T_c \leq T_L$. Superconductivity emerges from a fully formed heavy electron state. The general increase in $T_c$ with decreasing pressure is cut off by a competing state, quasiparticle localization, so $T_c$ reaches its maximum value at the pressure, $p_L$, at which the superconducting and localization transition lines intersect.

Region II: $T_c > T_L$ and $T_N$. Superconductivity emerges from a partially formed heavy electron state whose ability to superconduct is reduced by the partially hybridized local moments with which it coexists. The region includes the quantum critical point (QCP) at $T=0$ that marks a zero temperature transition from a state with partially localized ordered behavior to one that is fully itinerant[7]; this QCP is the origin of the quantum critical spin fluctuations that provide the pairing glue in all three regions[2].

Region III: $T_c < T_N$. Partially hybridized local moments are present in sufficient number to become antiferromagnetically ordered at the Néel temperature $T_N$ despite the presence of coexisting remnant heavy electrons that become superconducting at lower temperatures.

The dominance of superconductivity around the QCP supports the idea that the coupling of quantum critical spin fluctuations to the heavy electron quasiparticles plays a central role, with the resulting induced attractive quasiparticle interaction being maximally effective near it. Importantly, there is direct experimental evidence that these quantum critical fluctuations provide the superconducting glue. Curro et al.[4] found that at the pressure at which $T_c$ is maximum, the spin-lattice relaxation rate, $1/T_1$, to which these give rise, scales with $T_c$. A recent detailed investigation of that scaling[5] explains how this comes about: at this pressure $T_c$ scales with the coherence temperature, $T^*$, that marks the initial emergence of heavy electron behavior (and is given by the nearest-neighbor exchange interaction among the $f$-electron local moments[8]), while the scaling of $1/T_1$ with $T^*$ is a unique signature of its origin in quantum critical spin fluctuations.

In this communication we propose a simple BCS-like phenomenological expression for the superconducting transition temperature that incorporates these important scaling results, explains the variation of $T_c$ with pressure for both $CeCoIn_5$ and $CeRhIn_5$, and offers a detailed prediction for a similar dome-like structure in other quantum critical heavy electron superconductors.

We begin by recalling that for phonon-induced superconductivity, the BCS expression for $T_c$ depends on three quantities[6]: the quasiparticle density of states, the average strength, $V$, of the phonon-induced attractive interaction between quasiparticles, and the average energy range over which it is attractive. Our phenomenological heavy electron quantum critical magnetic expression involves analogues of the same three quantities and takes the form:

$$T_c(p) = 0.14 T_m^* \exp\left(-\frac{1}{N_F(p,T_c)V(p)}\right) = 0.14 T_m^* \exp\left(-\frac{1}{\eta\kappa(p)}\right), \quad (1)$$

where, as required by the above scaling results, we have taken the range of energies over which the quantum critical spin-fluctuation induced interaction will be attractive to be proportional to $T^*_m$, the coherence temperature at the pressure $p_L$, at which $T_c$ is maximum, while its prefactor, 0.14, can be determined from experiment as we show below. In equation (1), $N_F(p,T_c)$ is the heavy electron density of states at $T_c$ that, as discussed in Methods, is determined by the Region in which superconductivity emerges, we have made the physically plausible assumption that the effective heavy electron spin-fluctuation induced attraction, $V(p)$, scales with the pressure dependent interaction between local moments, $T^*(p)$, and is given by $V(p)=\eta k_B T^*(p)$, where $k_B$ is the Boltzmann constant and $\eta$ is a material-dependent parameter that measures the relative effectiveness, for a given material, of spin fluctuations in bringing about superconductivity, and we have introduced the dimensionless characteristic coupling strength, $\kappa(p)=N_F(p,T_c)k_B T^*(p)$.

It is important to note that since experiment shows that the heavy electron specific heat varies inversely as $T^*$ and grows logarithmically as the temperature is lowered[20,22], $C/T \sim 1/T^* \ln(T^*/T)$, the density of states, $N_F(p,T_c)$, will exhibit a similar dependence on $T^*(p)$, a quantity that experiment shows varies monotonically with increasing pressure (cf. the inset of Fig.2b); so *without a countervailing $T^*(p)$ dependence in the strength of spin-fluctuation induced interaction*, the dimensionless pairing strength would vary monotonically and equation (1) could never lead to the dome structure of $T_c$ seen experimentally.

Since equation (1) may be rewritten as

$$\ln\frac{T_c(p)}{T_m^*} = \ln 0.14 - \frac{\eta^{-1}}{\kappa(p)}, \qquad (2)$$

a plot of the experimental value of $\ln(T_c/T_m^*)$ against $1/\kappa(p)$ provides a test of our BCS-like expression for $T_c$. As discussed in Methods, in the absence of systematic specific heat measurements, $\kappa(p)$ may be determined from experiment by using a two-fluid analysis[17-22] to obtain the heavy electron density of states, $N_F$; the resulting values of $\kappa(p)$ for CeCoIn$_5$ and CeRhIn$_5$ are given in Fig. 2a. When used to test the validity of equation (2), we find, as may be seen in Fig. 2b, that the two materials fall on the same line, a scaling result that provides strong evidence for the validity of our BCS-like equation, while the common intercept tells us that $0.14T^*_m$ is the best choice for the range of the spin-fluctuation induced attraction for the two compounds.

Our model enables us to predict the maximum effectiveness of the spin-fluctuation induced interaction for a given material; it is given by

$$\lambda_{\max} = \eta \kappa(p_L) = \ln\left(0.14 T^*_m / T_c^{\max}\right)^{-1}. \qquad (3)$$

We further note that since $\kappa(p)$ is the only pressure dependent quantity in equation (1), our predicted ratio, $T_c(p)/T_c^{\max}$, involves no free parameters and takes the simple form

$$\frac{T_c(p)}{T_c^{\max}} = \exp\left[-\lambda_{\max}^{-1}\left(\frac{\kappa(p_L)}{\kappa(p)} - 1\right)\right]. \qquad (4)$$

As may be seen in Figs. 2c and 2d, when we use the values of $\kappa(p)$ shown in Fig. 2a as input and determine $\lambda_{\max}$ to be 1.23 for CeCoIn$_5$ and 0.62 for CeRhIn$_5$ from experiments at $p_L$, equation (4) provides a remarkably good quantitative explanation of the dome-like structure[4,5] observed as the pressure is varied in CeCoIn$_5$ and CeRhIn$_5$. We note that both $\kappa(p)$ and $T_c(p)$ are peaked at $p_L$, the pressure at which the delocalization line, $T_L$, intersects with $T_c$. Through the behavior of $N_F$ in the three regions, our model successfully explains the decrease in $T_c$ above this pressure as being brought about by the reduction in the heavy electron density of states produced by the increase in $T_L$; below this pressure, its decrease arises from the reduction in the heavy electron density of states brought about by the partial localization of the heavy electrons that reflects the incomplete hybridization of the *f*-moments.

Encouraged by the above results, we next apply our approach to the emergence of superconductivity in those heavy electron materials for which $T^*$ has been measured and Curro scaling has been established or appears likely to apply. Our results are given in Table I, where the characteristic dimensionless coupling strength, $\kappa(p_L)$, has been calculated using the two-fluid expression for $N_F$, and the effectiveness parameter, $\eta=\lambda_{max}/\kappa(p_L)$, is obtained at the measured or assumed optimal pressure, $p_L$. We call attention to a striking similarity in the values of $\lambda_{max}$ shown in Table I: $UPt_3$ appears to be a sister element to $CeCoIn_5$ and $PuCoIn_5$, even though their superconducting transition and coherence temperatures differ by a factor of five and their superconducting states possess different symmetries. The large value of $\eta$ found for $CeRhIn_5$ suggests that in this material the effective interaction, $V$, could be as large as $3T^*$, and the fact that $\eta>1$ for many materials suggests that the effective attractive interaction is generally somewhat greater than $T^*$.

Importantly, because there is only a modest variation in $\kappa(p_L)$ as one goes from one material to another, most of the measured variation in $[T_c/T^*]_{max}$ is likely due to variations in the impedance match between the spin-fluctuation spectrum and the heavy electron Fermi surface that we have parametrized by $\eta$. These variations can be explained by changes in effective dimensionality and crystal structure. As Monthoux and Lonzarich emphasized in their seminal papers[13,14]: near two dimensionality and a tetragonal crystal structure are most favorable to superconductivity; their presence in $CeRhIn_5$ at 2.4 GPa and $PuCoGa_5$ could explain the relatively large values of $\eta$ seen for these materials, while their absence in $CeIn_3$ would explain its low value of $\eta$ and its very low $T_c/T^*$.

$CeRhIn_5$ at 2.4 GPa and $PuCoGa_5$ demonstrate how very effective spin-fluctuations can be in bringing about superconductivity; their $T_c$ is an appreciable fraction of the effective heavy electron Fermi energy, $k_BT_c/E_F=2k_BT_cN_F(T_c)/3$, being 0.016 and 0.013 respectively, fractions large compared to those seen in the cuprates and very large to those found for conventional superconductors. Our model for heavy electron superconductivity leads to the prediction that the maximal value of the ratio $k_BT_c/E_F$ is ~0.03, about twice the above values.

As a first step toward understanding the microscopic origin of equation (1), we can ask whether it is consistent with the anticipated results of a microscopic strong coupling calculation of quantum critical spin-fluctuation induced superconductivity for heavy electron materials that takes full account of an experimentally-determined frequency dependent interaction. We find (see Methods) in the case of CeCoIn$_5$, where neutron scattering experiments yield direct information on the quantum critical spin fluctuation spectrum, that the range of the effective attractive interaction found in microscopic strong coupling calculations is remarkably close to that we propose phenomenologically, and complete consistency is obtained provided the coupling of quasiparticles to the spin-fluctuations scales with $T^*$. This is but a first step, but we hope that this consistency will further encourage the development of a complete microscopic derivation of our simple phenomenological BCS-like equation for $T_c$ in which quantum critical spin-fluctuation superconductivity can be characterized by a range, ~$0.14T^*_m$, and a pressure-dependent strength, $\eta T^*$, both of which can be determined directly from experiment. Another interesting questions for future study is whether our phenomenological approach to quantum critical spin-fluctuation induced superconductivity in heavy electron materials can be extended to the cuprates and any other unconventional superconductors in which scaling behavior for the spin-lattice relaxation rate with $T_c$ has been seen at or near optimal doping levels.

**Acknowledgements** Y.Y. is supported by the National Natural Science Foundation of China (NSFC Grant No. 11174339) and the Strategic Priority Research Program (B) of the Chinese Academy of Sciences (Grant No. XDB07020200). We thank G. Lonzarich for his critical reading and helpful comments on an earlier draft of this manuscript, Z. Fisk for his critical reading and most helpful remarks about the framing of this manuscript, and the Aspen Center for Physics (NSF Grant No. PHYS-1066293) and the Santa Fe Institute for their hospitality during its writing; Y.Y. thanks the Simons Foundation for its support.

**Author Contributions** Y.Y. analyzed the data; Y.Y. and D.P. developed the idea and wrote the manuscript.

Table I. $T_c$, $T^*_m$, and calculated parameters at the measured or proposed optimal pressure, $p_L$, for known quantum critical heavy electron superconductors.

|  | CeRhIn$_5$ | CeCoIn$_5$ | CeIrIn$_5$ | PuCoGa$_5$ | PuCoIn$_5$ | Ce$_2$PdIn$_8$ | Ce$_2$CoIn$_8$ | CeIn$_3$ | UPt$_3$ |
|---|---|---|---|---|---|---|---|---|---|
| $p_L$ (GPa) | 2.4 | 1.4 | 2.2 | 0 | 0 | 0 | 0 | 2.8 | 0 |
| $T_c(p_L)$ (K) | 2.3 | 2.6 | 1.05 | 18.5 | 2.5 | 0.7 | 0.4 | 0.2 | 0.5 |
| $T^*_m$ (K) | 37 | 92 | 100 | 430 | 100 | 25 | 50 | 80 | 20 |
| $T_c/T^*_m$ | 0.062 | 0.028 | 0.011 | 0.043 | 0.025 | 0.028 | 0.008 | 0.0025 | 0.025 |
| $\kappa(p_L)$ | 0.40 | 0.48 | 0.59 | 0.44 | 0.49 | 0.48 | 0.61 | 0.74 | 0.49 |
| $\lambda_{max}$ | 1.23 | 0.62 | 0.39 | 0.85 | 0.58 | 0.62 | 0.35 | 0.25 | 0.58 |
| $\eta$ | 3.09 | 1.30 | 0.66 | 1.94 | 1.18 | 1.29 | 0.57 | 0.34 | 1.18 |
| $k_B T_c/E_F$ | 0.016 | 0.009 | 0.004 | 0.013 | 0.008 | 0.009 | 0.003 | 0.001 | 0.008 |
| *refs.* | 7,18 | 11,19 | 24,25 | 3,4 | 26 | 27 | 28 | 29 | 8,30 |

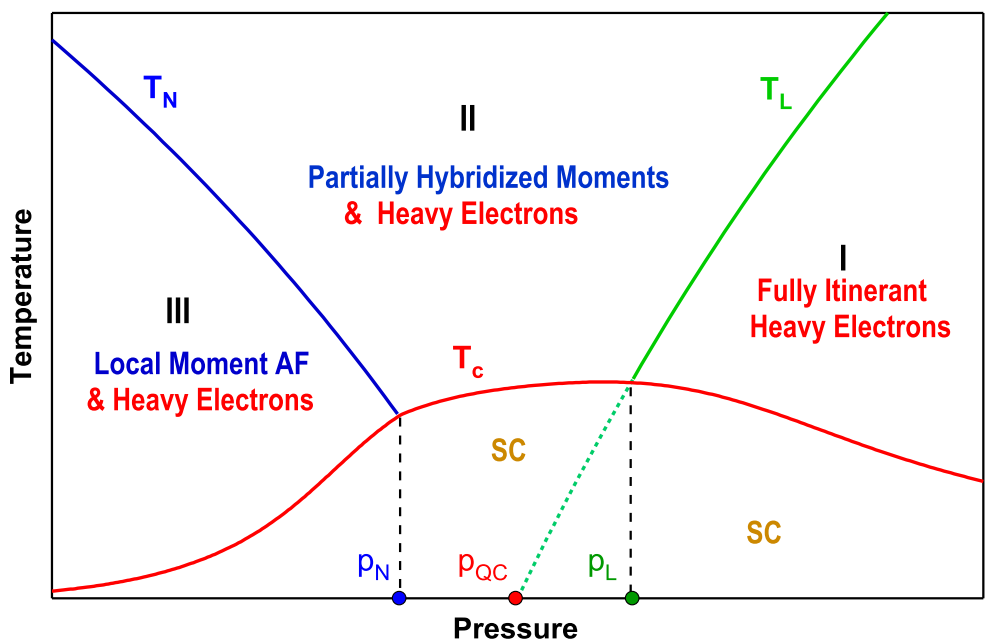

Figure 1. **A phase diagram for heavy electron superconductors**[5,7,18,19]. In region I, only itinerant heavy electrons exist below $T_L$ due to complete hybridization of the *f*-moments with background conduction electrons; in region II, collective hybridization is not complete so that heavy electrons coexist with partially hybridized local moments; in region III, these residual moments order antiferromagnetically (AF) at $T_N$ and the surviving heavy electrons become superconducting (SC) at a lower temperature, $T_c$. The coupling of heavy electrons to the magnetic spin fluctuations emanating from the QCP is responsible for the superconductivity in all regions.

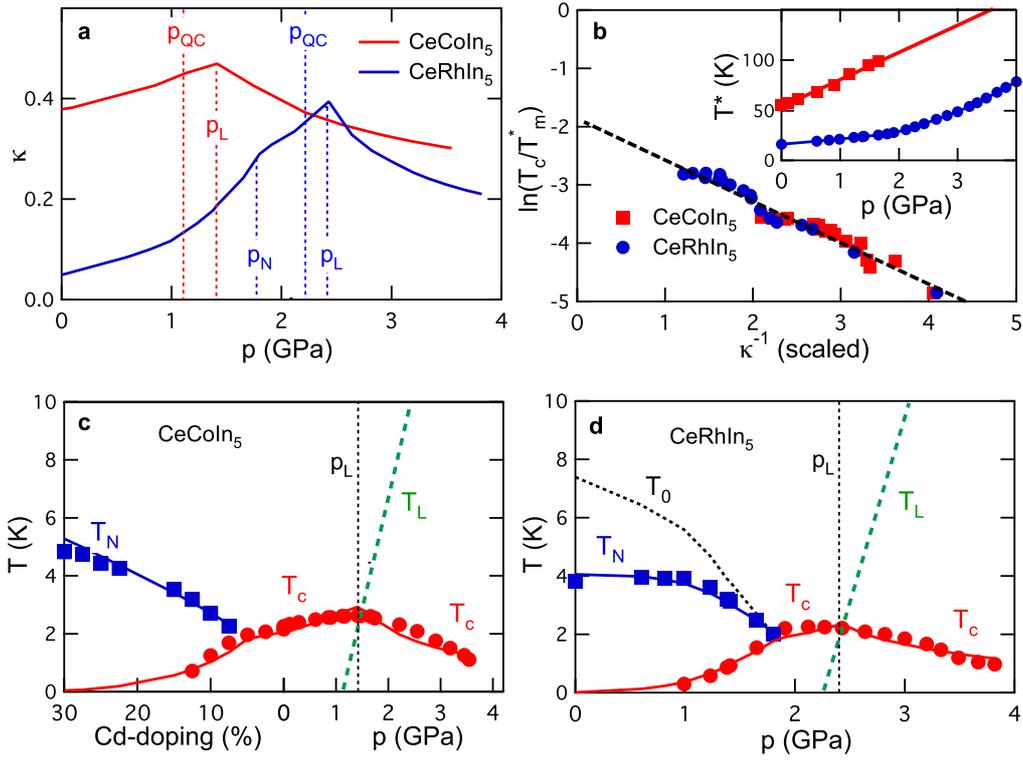

Figure 2. **Comparison of theory and experiment for the ordering temperatures measured in CeCo(In$_{1-x}$Cd$_x$)$_5$ and CeRhIn$_5$. a,** Pressure variation of the predicted dimensionless pairing strength (see Methods), $\kappa(p)=k_B T^*(p) N_F(p,T_c)$. **b,** Scaling of $\ln(T_c/T^*_m)$ and $\kappa(p)^{-1}$ (scaled) for CeCoIn$_5$ and CeRhIn$_5$. The inset shows the experimental values of $T^*(p)$ that are used to obtain $\kappa(p)$ in both compounds[7,19]. **c,** Comparison of the predicted (solid lines) and experimental[19,21] $T_c$ and $T_N$ in CeCo(In$_{1-x}$Cd$_x$)$_5$ and CeCoIn5 with $\eta$=1.30 and $\lambda_{max}$=0.62. **d,** Comparison of the predicted (solid lines) and experimental[18] $T_c$ and $T_N$ in CeRhIn$_5$ with $\eta$=3.09 and $\lambda_{max}$=1.23.

**Methods**

**1. Determining the characteristic dimensionless pairing strength, $\kappa(p)$, from experiment**

In the Fermi liquid regime (Region I in Fig. 1), where the density of states can be derived from the specific heat measurements and the coherence temperature, $T^*(p)$, can be estimated from the resistivity, the dimensionless pairing strength, $\kappa(p)$, can be directly determined from experiment, so that our proposed BCS-like equation (4) involves no free parameters and could be verified without any further assumptions. However, because the relevant experimental information on the pressure dependence of the specific heat is not yet generally available, to test the applicability of equation (4) to heavy electron materials under pressure, we have used the two-fluid model to determine the pressure dependence of the density of states. This procedure has earlier been shown to yield correct specific heat results for a number of heavy electron compounds[7].

**1.1. The delocalization line, Néel temperature, and heavy electron density of states**

In the two-fluid model, the three regions in Fig. 1 are determined by the hybridization parameter[17],

$$f(p,T) = f_0(p)\left(1 - T/T^*(p)\right)^{3/2}, \qquad (5)$$

that quantifies the fraction of $f$-electrons that become itinerant. The pressure dependence of the hybridization effectiveness, $f_0(p)$, discussed below, can be determined from magnetic experiments (cf. ref. 11). For $f_0 > 1$, the line of complete delocalization temperatures, $T_L$, in the heavy electron phase diagram is obtained by setting $f(T_L)=1$, so that

$$T_L(p) = T^*(p)\left[1 - f_0(p)^{-3/2}\right]. \qquad (6)$$

For $f_0<1$, a fraction of residual local moments always remains and becomes antiferromagnetically ordered at low temperatures. The two-fluid model predicts that the Néel temperature, $T_N$, is given by[7]

$$\frac{T_N(p)}{T^*(p)} = \eta_N\left[1-f(T_N,p)\right], \quad (7)$$

where the frustration parameter, $\eta_N$, is independent of pressure and found to be 0.14 for CeCoIn$_5$ and 0.32 for CeRhIn$_5$ in Figs. 2c and 2d.

The heavy electron density of states in equation (1) is obtained in the two-fluid model by assuming that $N_F(T)$ follows the heavy electron specific heat[7,10,12], according to $N_F(T) = \frac{3}{\pi^2 k_B^2} f(T) \frac{C_{HE}}{T}$; experiment shows that the latter grows logarithmically as the temperature is lowered[10,12], $\frac{C_{HE}}{T} = \frac{k_B \ln 2}{2T^*}\left(1+\ln\frac{T^*}{T}\right)$, where the prefactor in $C_{HE}/T$ is determined by requiring the entropy at $T^*$, $S(T^*) = \int_0^{T^*} dT \frac{C_{HE}}{T} = k_B \ln 2$ and is cancelled out in our proposed equation (4). The logarithmic growth is cut off by complete delocalization at $T_L$ in Region I, superconductivity at $T_c$ in Region II, and long-range magnetic order at $T_N$, or its precursor at $T_0$ in region III so that the heavy electron density of states at the superconducting transition at $T_c$ is:

$$N_F(p,T_c) = \frac{3\ln 2}{2\pi^2 k_B T^*(p)} f_0(p)\left(1-\frac{T_x(p)}{T^*(p)}\right)^{3/2}\left(1+\ln\frac{T^*(p)}{T_x(p)}\right), \quad (8)$$

where $T_x(p)=T_L(p)$ in Region I, $T_c(p)$ in Region II and $T_{0/N}(p)$ in Region III (see Figs. 2c and 2d). Importantly, we see that because $N_F$ varies inversely with $T^*$, the characteristic dimensionless pairing strength, $\kappa(p)=k_B T^*(p)N_F(p,T_c)$, depends comparatively weakly on $T_x/T^*$ in all three regions:

$$\kappa(p) = \frac{3\ln 2}{2\pi^2}\left(1+\ln\frac{T^*(p)}{T_L(p)}\right), \qquad \text{(Region I)} \qquad (9a)$$

$$\kappa(p) = \frac{3\ln 2}{2\pi^2}f_0(p)\left(1-\frac{T_c(p)}{T^*(p)}\right)^{3/2}\left(1+\ln\frac{T^*(p)}{T_c(p)}\right), \qquad \text{(Region II)} \qquad (9b)$$

$$\kappa(p) = \frac{3\ln 2}{2\pi^2}f_0(p)\left(1-\frac{T_0(p)}{T^*(p)}\right)^{3/2}\left(1+\ln\frac{T^*(p)}{T_0(p)}\right). \qquad \text{(Region III)} \qquad (9c)$$

Because $f(p_L, T_c^{max})=1$, we find a simple formula for the maximal value of $\kappa(p)$ at $p_L$:

$$\kappa(p_L) = \frac{3\ln 2}{2\pi^2}\left(1+\ln\frac{T_m^*}{T_c^{max}}\right). \qquad (10)$$

**1.2. Deducing other key parameters from experiment**

The pressure dependence of the coherence temperature, $T^*(p)$, may be obtained from resistivity measurements[7,19], and, in the case of Cd-doped CeCoIn$_5$, from Knight shift experiments[22]. The results are shown in the inset of Fig. 2b.

To determine $f_0(p)$, we first note $f_0(p_{QC})=1$, and use experiment to determine $f_0$ at ambient pressure; for other pressures, we assume that $f_0(p)$ scales linearly with $T^*(p)$ (cf. ref. 11) and obtain

$$f_0(p) = 1 + (1-f_0(0))\frac{T^*(p)-T^*_{QC}}{T^*_{QC}-T^*(0)}, \qquad (11)$$

where $f_0(0)$ is the hybridization parameter at ambient pressure and $T^*(0)$ and $T^*_{QC}$ are the coherence temperatures at ambient pressure and the QCP, respectively.

For CeRhIn$_5$, one has $p_{QC} \sim 2.25$ GPa and $T^*_{QC} \sim 33$ K (ref. 18); an analysis of its magnetic properties yields $T^*(0) \sim 17$ K and $f_0(0) \sim 0.65$ at ambient pressure. For CeCoIn$_5$, a scaling analysis of the resistivity[20] suggests $p_{QC} \sim 1.1$ GPa and $T^*_{QC} \sim 82$ K, while an analysis of the temperature-magnetic field phase diagram yields $f_0(0) \sim 0.87$ and $T^*(0) \sim 56$ K at ambient pressure, a result that yields an excellent fit to the variation of the QCP with pressure[11]. For Cd-doping, we assume that 5% Cd-doping has similar effect on $f_0$ as a negative pressure of -0.7 GPa, as is suggested by experiment[21]. The effect of Cd-doping is, however, different from pressurization since $T^*$ is doping independent as is seen in the nuclear magnetic resonance experiment[22]. For both materials, our choice of $f_0(p)$ leads to a unique prediction of $T_L(p)$ that can be verified experimentally.

The cut-off temperatures, $T_x(p)$, for the growth in the heavy electron state density in Region III are determined from the Knight shift and/or Hall measurements[7]. For $CeCo(In_{1-x}Cd_x)_5$, experiment shows that $T_x$ is roughly given by $T_N$; for $CeRhIn_5$, experiment[7] shows that $T_x = T_0 \sim 2T_N$ at ambient pressure and decreases to $T_N$ at $p_N \sim 1.8$ GPa. In this region, a further experimental test of our choice of parameters is provided by the Néel temperature that can be calculated using equation (7).

On combining and inserting these experimental parameters into equation (9), we obtain the dimensionless pairing strength, $\kappa(p)$, in Fig. 2a, and the results for $T_N$ and $T_c$ shown in Figs. 2c and 2d that are in remarkably good agreement with experiment.

## 2. Prediction of a dome-like structure for $T_c$

Our prediction of a dome-like structure for $T_c$ versus pressure for any heavy electron superconductor is based on the behavior of the solutions of equation (4) for the three distinct regions of emergent superconductivity:

Region I: $f_0 > 1$ and $T_c < T_L$. The growth of $N_F(T)$ is cut off at the delocalization temperature, $T_L$, below which $f(T) = 1$ and equation (4) only depends on $f_0$,

$$\frac{T_c}{T_c^{max}} = \exp\left[-\lambda_{max}^{-1}\left(\frac{1+\ln(T_m^*/T_c^{max})}{1-\ln(1-f_0^{-2/3})}-1\right)\right]. \qquad (12a)$$

$T_c$ is maximum at the pressure at which $T_c=T_L$; it decreases at higher pressures because the density of states decreases, being cut off at higher values of $T_L$ by the increase in $f_0$.

Region II: $f_0\sim 1$ and $T_c>T_L$ and $T_N$. Because the growth of $N_F(T)$ extends to $T_c$, equation (4) takes the form,

$$\frac{T_c}{T_c^{max}} = \exp\left[-\lambda_{max}^{-1}\left(\frac{1+\ln(T_m^*/T_c^{max})}{f_0(1-T_c/T^*)^{3/2}(1+\ln(T^*/T_c))}-1\right)\right], \qquad (12b)$$

and has to be solved self-consistently. Most heavy electron quantum critical superconductors fall in this region, where the logarithmically nearly divergent density of states acts to enhance the effective interaction by a factor, $[1+\ln(T^*/T_c)]$, that can vary between 7.0 and 3.8 as one goes from $T_c/T^*=0.0025$ to 0.062.

Region III: $f_0<1$ and $T_c<T_N$. The growth in $N_F$ is cut off at $T_0$ so that

$$\frac{T_c}{T_c^{max}} = \exp\left[-\lambda_{max}^{-1}\left(\frac{1+\ln(T_m^*/T_c^{max})}{f_0(1-T_0/T^*)^{3/2}(1+\ln(T^*/T_0))}-1\right)\right]. \qquad (12c)$$

With increasing pressure, $f_0$ increases and $T_N$ and $T_0$ decrease, so that $T_c$ increases and becomes greater than $T_N$ before one reaches the quantum critical pressure.

## 3. A consistency check with microscopic strong coupling calculations

It is reasonable to assume that the pairing interaction for heavy electron superconductivity is given by an expression identical to that used to explain quantum critical cuprate superconductivity[2],

$$V(\mathbf{q},\omega) = g^2\chi(\mathbf{q},\omega). \qquad (13)$$

where $g$ is the quasiparticle-spin fluctuation coupling strength and $\chi(\mathbf{q},\omega)$, the dynamic susceptibility, follows the quantum critical form expected from its proximity to an antiferromagnetic state[13]

$$\chi(\mathbf{q},\omega) = \frac{\chi_\mathbf{Q}}{1+(\mathbf{q}-\mathbf{Q})^2 \xi^2 - i\omega/\omega_{SF}}, \qquad (14)$$

with a peak at the ordering wave vector, $\mathbf{Q}$, of magnitude $\chi_\mathbf{Q} = \pi\chi_0 (\xi/a)^2$ where $\xi$ is the antiferromagnetic correlation length, $a$ is the lattice constant, and $\chi_0$ is the uniform spin susceptibility, and a temperature-dependent spin fluctuation energy, $\omega_{SF}$. Since the measured ratios of the energy gap to $T_c$ for heavy electron materials are typically large compared to the weak coupling result, 1.75, any attempt to seek consistency between our proposed phenomenological expression for $T_c$ and microscopic calculations should begin with the strong coupling numerical results[14,15] required to take account of the frequency dependence of the interaction, equation (13). Although these have yet to be carried out for heavy electron materials, it is to be expected that these will yield a BCS-like expression in the strong coupling limit that is analogous to that found for the cuprates, namely,

$$T_c = \lambda_1 \omega_{SF} (\xi/a)^2 \exp\left(-\frac{1}{\lambda_2 g N_F(T_c)}\right), \qquad (15)$$

where $\lambda_1$ and $\lambda_2$ are constants of order unity.

The microscopic result, equation (15), will be consistent with our phenomenological expression, equation (1), if, first, the proposed microscopic prefactor, $\lambda_1 \omega_{SF}(\xi/a)^2$, is identical to $0.14 T^*_m$, the effective range over which we have proposed that the quantum critical spin fluctuation induced interaction will be attractive, and second, if the coupling, $g$, of the heavy electron quasiparticles to the spin fluctuations scales with $T^*$, the nearest neighbor local moment interaction[8]. This last connection is plausible since through collective hybridization the heavy electron quasiparticles are born coupled by an interaction similar to that of the local moments from which they emerge. Importantly, experimental information on the microscopic prefactor is available for CeCoIn$_5$, where neutron scattering measurements of the spin fluctuation spectrum near $T_c$ at ambient pressure[23] yield $\omega_{SF}$=0.3±0.15 meV and $\xi$=9.6±1.0 Å (about twice the in-plane lattice constant $a$=4.60 Å). One then has $\omega_{SF}(\xi/a)^2$=1.3 meV~15.1 K, in remarkably close agreement with our phenomenological result, $0.14T^*_m$=12.9 K. The two expressions agree if we take $\lambda_1$=0.85 in equation (15) and assume that neutron scattering at the quantum critical pressure will yield results for this product that are similar to those found at ambient pressure. Future calculations and experiments on other materials can test our prediction that the microscopic prefactor will always be ~0.14 $T^*_m$.